\documentclass{ar-1col}

\usepackage[numbers,compress,square]{natbib}
\usepackage{url}
\usepackage{subfigure}
\usepackage{graphicx}
\usepackage{amsmath,amssymb}
\usepackage{aas_macro}

\makeatletter
\renewcommand{\@thesubfigure}{\hskip\subfiglabelskip}
\makeatother
\newcommand{\gray}{$\gamma$-ray~}
\newcommand{\grays}{$\gamma$-rays~}
\jname{Annu.Rev.Nucl.Part.Sci}
\jvol{73}
\jyear{2022}
\doi{10.1146/((please add article doi))}

\begin{document}

% Page header
\markboth{Z. Cao et al.}{UHE Gamma-Ray Astronomy}

% Title
\title{Ultra High Energy Gamma Ray Astronomy}

%Authors, affiliations address.
\author{Zhen Cao,$^{1,2,3}$ Songzhan Chen,$^{1,2,3}$ Ruoyu Liu,$^{4,5}$ and Ruizhi Yang$^6$
\affil{$^1$Key Laboratory of Particle Astrophysics, Institute of High Energy Physics, Beijing, China, 100049; email: caozh@ihep.ac.cn}
\affil{$^2$Physics Department, University of Chinese Academy of Sciences, Beijing, China, 100049}
\affil{$^3$Tianfu Cosmic Ray Research Center, Chengdu, China, 610000}
\affil{$^4$School of Astronomy and Space Science, Nanjing University, 210023 Nanjing, Jiangsu, China}
\affil{$^5$Key laboratory of Modern Astronomy and Astrophysics (Nanjing University), Ministry of Education, Nanjing 210023, China}
\affil{$^6$University of Science and Technology of China, 230026 Hefei, Anhui, China}}
%Abstract
\begin{abstract}
Ultra-High Energy (UHE, $>$0.1\,PeV) $\gamma$-ray Astronomy is rapidly evolving into an expanding branch of the $\gamma$-ray astronomy with the surprising discovery of 12 PeVatrons and the detection of a handful of photons above 1 PeV.
%It has been observed that 
Nearly all known celestial object types that have emissions in the TeV band are found also emitting UHE photons. UHE $\gamma$-rays have a well-defined horizon inside our galaxy due to the absorption of infrared and cosmic microwave backgrounds in the universe. With the last 30 years, traditional cosmic ray (CR) detection techniques allow the detection of UHE $\gamma$-rays, and opened up the last observation window. For leptonic sources, UHE radiation is in the deep Klein-Nishina regime which is largely suppressed. Therefore UHE $\gamma$-ray detection will help to locate and identify hadronic radiation sources, tracing the historic pursuit for the origin of CRs around the knee of the spectrum. The Crab Nebula is again the focus of attention with measured photon emissions above 1\,PeV. In the absence of hadronic processes, this may indicate the existence of an extreme accelerator of e$^+$/e$^-$.  Utilization of the CR extensive air shower detection techniques broadens the field of view of the source observations, enabling the measurement of UHE radiation surrounding the sources. These observations can probe the particle propagation inside and outside the accelerators and the subsequent injection/escape into the interstellar medium. 
%The detection of UHE photons with the unprecedented sensitivity also provide opportunities to test fundamental laws of physics and explore new phenomena. 
\end{abstract}

%Keywords, etc.
\maketitle

%Table of Contents
\tableofcontents

% Heading 1
\section{INTRODUCTION}    
Very-High Energy (VHE) $\gamma$-ray astronomy has been experiencing enormous improvement in understanding the non-thermal universe in the past three decades\citep{Hinton_2009}. From the detection of the first TeV photon from the Crab Nebula \citep{Whipple_Crab}, not only the number of sources that emit VHE $\gamma$-rays above 0.1 Terra electron Volt (TeV, 10$^{12}$ eV)  grows exponentially with time but also the types of astrophysical objects as the candidates of the VHE $\gamma$-ray emitters. New phenomena and radiation mechanisms constantly push the field forward to new territory with milestone discoveries\citep{Funk_2015}. These discoveries make VHE $\gamma$-ray astronomy the most productive and successful sub-field in the high-energy domain.
This paper focuses on photons with even higher energy, around 1 Peta-electronvolt (PeV, 10$^{15}$ eV); photons with energy above 0.1 PeV are dubbed as Ultra-High Energy (UHE) photons.
Historically, many indications of photons above 1 PeV \citep{Kiel_Cygnus-X3} drove a wave of development of $\gamma$-ray detection based on cosmic ray (CR) extensive air shower measurement techniques. Detecting photons at 1 PeV provides the most direct  evidence of the parent charged particles around 10 PeV in the sources. The acceleration mechanisms of these high energy particles remain unclear after the discovery of CR more than a century ago. 

After 30 years of development \citep{Cao_2021-Univ}, techniques to detect UHE photons have matured with a sensitivity up to 10$^{-14}\,$TeV\,cm$^{-2}$s$^{-1}$ at 0.1 PeV and effectively measure the emissions from many known VHE $\gamma$-ray sources, including the standard candle in the VHE domain, the Crab Nebula. Ushering in the era of UHE astronomy, the Large High Altitude Air Shower Observatory (LHAASO)\citep{Cao_2021-NA} rapidly discovered a dozen sources\citep{lhaaso_nature} with a stable flux of UHE photons. Many of these sources have power-law-like spectral energy distributions (SED) without a clear cut-off feature. The long-standing assumptions of the upper limits of particle acceleration within Galactic sources were well below 1 PeV. These discoveries have reset the upper limit to a much higher value, opening a broader territory of the non-thermal regime where there are many candidates of origins of cosmic rays above 1 PeV, defined as PeVatrons. New sources with unknown features, such as $\gamma$-ray morphological structures, have extended the field to explore new radiation mechanisms and particle acceleration procedures. 12 UHE $\gamma$-ray sources brighter than 0.7 Crab Unit (CU, the flux from the Crab Nebula) are observed in the northern hemisphere using LHAASO,  designed with a sensitivity of 12 milli-CU. They are conceivably associated with all types of known candidates, such as Supernova Remnants (SNRs), Pulsar Wind Nebulae (PWNe), Young Massive-star Clusters (YMCs), and micro-quasars. These sources currently have been linked to VHE $\gamma$-ray emissions. More comprehensive surveys in the upcoming years will unveil new sources. In parallel,  detailed investigations into known sources will continue.

This paper is arranged as follows. The second section is devoted to describing the domain of the UHE $\gamma$-ray astronomy with a natural horizon due to the absorption of cosmic microwave background (CMB) and cosmic infrared background. The third section describes the development of the instruments and the critical technology, distinguishing between air showers induced by $\gamma$-rays and protons. The fourth section introduces the discovery of PeVatrons and possible candidate astrophysical object sources. The fifth section explains the deep investigation of the Crab Nebula, the best-studied object for its radiation and particle acceleration in the UHE domain, focusing on the possible extreme acceleration of electrons/positrons. The sixth section discusses the most favourable candidates of CR factories, including the Cygnus region and SNR~G106.3+2.7. Pulsar halos, a relatively new topic in spatially extended sources, are reviewed in the seventh section for both phenomenological and observational studies. The eighth section introduces the recent efforts of measuring diffuse UHE $\gamma$-ray emission from the Galactic Plane and their implications. The ninth section is a summary of the review.

%Heading 1
\section{THE HIGHEST ENERGY BAND OF ELECTROMAGNETIC OBSERVATION OF THE UNIVERSE}
This section discusses the absorption of UHE photons through photon-photon interactions, such as CMB and infrared photons in the universe, creating a low-energy photon background. Therefore, the UHE domain has a well-defined horizon. We examine the source of the UHE photons, notably the PeVatrons, their definition, possible candidates, and distribution in the universe. We also discuss PeVatrons as the origin of cosmic rays and their relationship with the knee of the CR spectrum. 

% Heading 2
\subsection{Absorption of Gamma-rays in the Path to the Earth}

High energy $\gamma$-rays interact with the background photo fields inevitably via pair productions ($\gamma \gamma \rightarrow e^+e^-$).  In this process \grays are attenuated. CMB, the interstellar radiation fields (ISRF) \citep{moskalenko06,Popescu17} in our Galaxy  and the extragalactic background light (EBL) \citep{franceschini08} contribute to the background photon fields. 
The photon-photon pair-production cross-section averaged over directions of the background-radiation field
depends on the product of energies of colliding photons. The energy dependence of the pair production cross section is given by Gould \& Schr{\'e}der \cite{gould66}. For the given energy of the $\gamma$-ray photon 
$E_\gamma$, it peaks at the  wavelength of background photons $\lambda \sim 2.5 (E_{\gamma}/{\rm 1 \ TeV}) \mu \rm m$. CMB is characterised as black body radiation with a temperature of about $2.7~\rm K$ and its SED  peaks at $\lambda \sim 1~ \rm mm$, while EBL and ISRF are mainly contributed by the emission from dust whose temperature is less than  $100~\rm K$ and their SEDs peak at $\lambda \sim 100~ \rm \mu m$.  In the local universe, CMB starts to dominate the $\gamma$-ray opacity above 100\,TeV, while EBL dominates at lower energies.
The opacity has been calculated by performing the line-of-sight integral of the product of the pair production cross section with the energy density of the radiation fields. The $\gamma$-ray opacity at 1~PeV is already larger than unity when the distance of the source larger than 10~kpc, which means that we can hardly detect the PeV photons with extragalactic origin. For 100 TeV photons, the mean-free path (at which distance the opacity is equal to 1) in typical EBL is estimated as 1.5~Mpc \citep{Popescu17}. Inside our Galaxy the ISRF also contribute significantly to the $\gamma$-ray opacity. Since ISRF, unlike EBL and CMB, is highly inhomegeneously distributed in our Galaxy, the $\gamma$-ray opacity also depends strongly on the direction of sources. If the line of sight of a source passes through the Galactic Center (GC), the effect of ISRF on 100 TeV $\gamma$-ray opacity  approaches that of the EBL for the source at a distance of 1~Mpc, and the derived opacity is 0.7. In such a case, about a half of the 100 TeV photons will be absorbed by ISRF. We note that along this direction a 100 TeV photon receives the largest possible attenuation   in our Galaxy and the opacity drop significantly as the increasing latitude and longitude of the line of sight.   At even lower energy, the opacity also drops sharply, because the corresponding background photons below 100 TeV is dominated by the Wien side of the dust thermal emissions, whose number density drop significantly as the  wavelength decreases (increasing background photon energy). At about 20 TeV, the ISRF is already transparent for $\gamma$-rays. 

In conclusion, CMB dominates the opacity for PeV photons, and limits the horizon of PeV photons to our Galaxy. At 100 TeV, $\gamma$-ray sources are visible in the Local Group ($\sim 1~\rm Mpc$), and the $\gamma$-rays from Galactic sources are only marginally attenuated if they locate towards GC.  Thus above $100~\rm TeV$ is a suitable window for Galactic astronomy. 

%The dependence of the pair production cross section on the energy is given in Ref.\citet{Gould66}.

\subsection{The Window of Search for Galactic PeVatrons}

The `knee', a break in the energy spectrum of CRs measured at the Earth around 1 PeV  ($10^{\rm 15}~\rm eV$)\citep[see, e.g., ][]{argo_knee}, is a significant feature. The current paradigm of CRs also postulate that at least to PeV energy the CRs should have a Galactic origin \citep[see, e.g., ][]{Blasi13}. Thus one of the key issue in CR science is to identify the PeV particle accelerators, which are dubbed as PeVatrons,  in our Galaxy. 

CRs are charged particles and will be deflected by the Galactic magnetic field. As a rule of thumb estimation, the Larmor radius $r_L$ can be estimated:
\begin{equation}
r_L\sim 10^{12} (E_p/1~\rm GeV) (B/3~\rm \mu G)^{-1}~\rm cm 
\end{equation}
where $E_p$ is the energy of the relativistic proton, $B$ is the magnetic field.
The magnetic field strength in our Galaxy lie in the range $1-10~\rm \mu G$, with an average value of $3~\rm \mu G $ in the Galactic disk.
Even for protons at energies as high as 1 PeV,  $R_L$ is only as small as $\sim$1 pc assuming a  magnetic field of 3$\mu G$, which is much less than the distance to any possible CR source. As a result, the anisotropy  in CR arrival direction measurements cannot provide decisive information on the CR sources. On the other hand, $\gamma$-rays, as the secondary production of CRs interacted with ambient gas, propagate rectilinearly and can be used to trace the CR acceleration sources. The $\gamma$-ray carries about $1/10$ of energy of the parent CR's \citep{kelner06}, therefore PeV CR protons are expected to produce  $\gamma$-rays at energies $\sim 100~\rm TeV$, which is the UHE domain. 

Supernova remnants (SNRs) are regarded as the most promising CR accelerators in our Galaxy. GeV \gray observations have already found the pion-decay feature of the \gray emissions from Mid-aged SNRs \citep{agile_w28,fermi_pion}, which are regarded as a strong proof that these Mid-aged SNRs do accelerate CR protons. However, every star like the Sun can generate particles up to energies above 10 GeV. The question of whether SNRs can account for the CRs up to PeV is still open.  The mid-aged SNRs cannot be PeVatrons because the observed \gray spectrum reveal cutoff at dozens of GeV, which corresponds to a cutoff in parent proton spectrum around several hundred GeV. The younger SNRs are indeed TeV \gray emitters\citep{hess_1006,hess_1713,veritas_tycho,casa_magic}, but the production mechanism of these TeV \grays remains still unclear. In this energy range the \gray production mechanism in the astrophysical process are inverse Compton scattering (IC) of relativistic electrons off low energy background photon fields and neutral Pion decay process in the inelastic scattering of CR nuclei with ambient gas. Above the produced \gray energy of $\sim 100~\rm TeV$, the IC processes go into deep Klein-Nishina regime even for the CMB as the low energy photon fields, the produced \gray spectra will be softened inevitably in this energy range \citep{hess_1641}.  Therefore, before LHAASO's operation, a hard spectrum above $\sim 100~\rm TeV $ without a significant softening can only be formed in pion-decay process with parent CR proton energy larger than several hundred TeV and can be regarded as a strong hint of hadronic PeVatrons. 

The above approach  has been firstly applied by H.E.S.S collaborations in HESS J1641-463 \citep{hess_1641}, in which a hard \gray spectrum with a spectral index of $-2$ extending to about $20~\rm TeV$ is detected. Such a spectrum can be explained with IC process only if the cutoff of electron spectrum exceed $700~\rm TeV$, which is extremely difficult in the corresponding SNR environment. On the other hand, the pion-decay process is a more natural explanation and the observed \gray spectrum sets the $99\%$ confidence level lower limit of the parent proton spectrum to be $100~\rm TeV$. Clearly, a high precision \gray spectral measurements in even higher UHE domain will reveal the origins of the PeVatrons.   

The systematic way to identify PeVatrons thus reveals the origin of high energy CRs requires high precision \gray spectral measurements in even higher energies, i.e, in the UHE domain.

\subsection{The Origin of Cosmic Rays above the Knee}
The origin of CRs above the knee ($10^{15}~\rm eV$) is still unknown and even more mystery than that for CRs below the knee.
For  CRs around $10^{18} ~\rm eV$, the gyro-radius, about $400~\rm pc$ according to Equation 1, is comparable to the thickness of the Galactic disk.
In this energy regime, two features have been identified in the CR spectrum, namely the `second knee’ around $10^{17.5} ~\rm eV$, where the chemical composition changes
significantly as measured by HiRes\citep{hires-2nd-knee}, and the following `dip’ in
the spectrum \citep{berezinsky06} or the `ankle’ around $10^{18.5}~\rm eV$ where the CR spectrum becomes flatter. The second knee is a significant break in CR spectrum,  at which the spectrum steepens  from index -3.0 to approximately -3.3.  It is widely accepted to be the upper limit of Galactic CR accelerators, such as SNRs, but could also be consistent with the hypothesis that the CRs at higher energies escape more freely from the Galaxy (e.g., Ref.\cite{grenier15}).
The spectrum hardening at the ankle is widely accepted to be the indication of the onset of the   extragalactic component.  What is the origin  of CRs between the knee and ankle? Or, which portion of those CRs are accelerated by Galactic sources? What type of sources are responsible? All the questions are  widely open, although the exact energy of the knee is still unclear because of the uncertainties of the chemical composition of CRs in this energy range. The UHE \gray observations, particularly the collections of photons above 1\,PeV from different sources, seem quite promising to pin down those problems. Firstly, the direct \gray observations can directly measure the maximum acceleration energies. Secondly, the diffuse \gray emissions from the Galactic plane and nearby giant molecular clouds (GMCs) provide independent measurements of CR spectra and chemical composition, since the \gray spectrum depends on kinetic energy per nucleon, rather than the total kinetic energy of the nucleus, different chemical composition will produce different \gray spectrum. In conclusion, UHE \grays can provide unique and important information on the knee and CR origins above the knee.

\section{INSTRUMENTS OF UHE GAMMA-RAY ASTRONOMY}
In this section, we introduce existing instruments, ongoing projects, and impending detectors. Taking the historical point of view, we discuss the critical issue of CR background suppressing techniques and how the extensive air shower (EAS) techniques developed into a successful $\gamma$-ray detection tool. Depending on the FoV of the instruments, they are used for two primary goals: surveying a large number of sources and targeted observations for in-depth investigations in radiation mechanisms and particle accelerations in sources.
 
\subsection{Historical and Technical Remarks about EAS Technique in $\gamma$-ray Astronomy}
\begin{figure}[!h]
%\captionsetup[subfigure]{labelformat=empty}
\subfigure[]{
\begin{minipage}[b]{0.58\textwidth}
\includegraphics[width=1.0\textwidth]{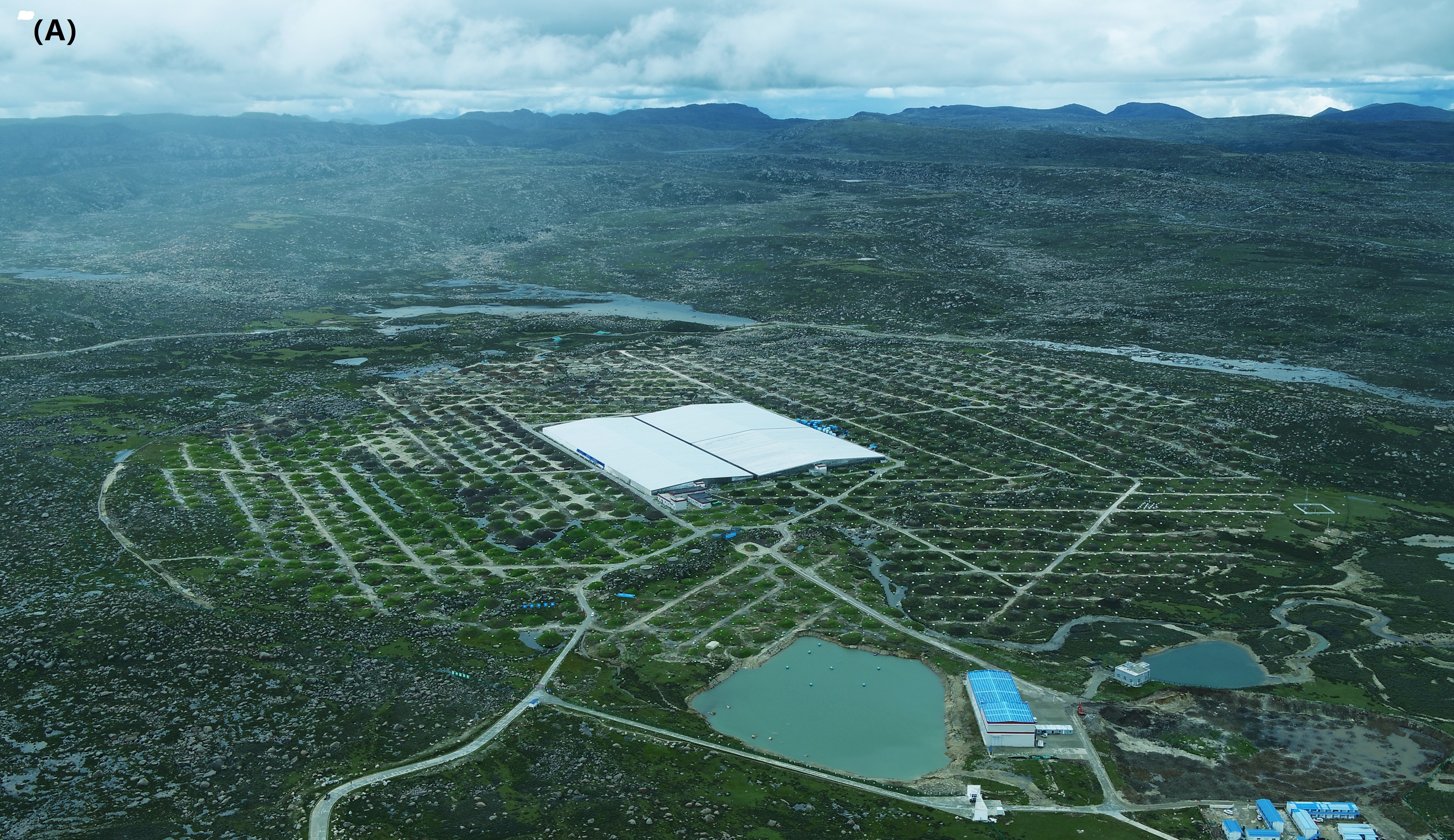}
\end{minipage}
}
\subfigure[]{
\centering
\begin{minipage}[b]{0.37\textwidth}
\includegraphics[width=1\textwidth]{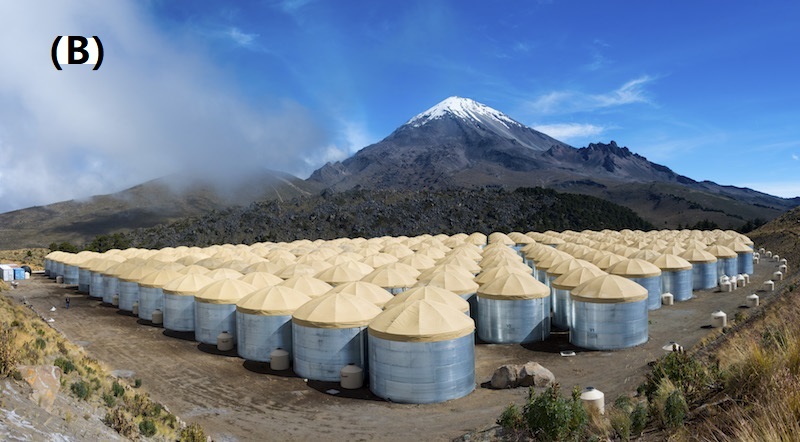}
\includegraphics[width=1\textwidth]{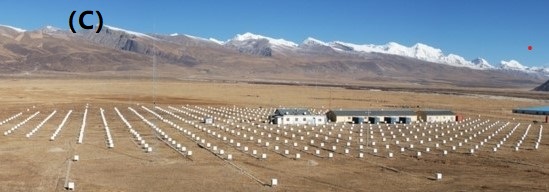}
\end{minipage}
}

\caption{The current three experiments in UHE $\gamma$-ray astronomy. They are all using cosmic ray extensive air shower detecting techniques. The panel (A) is the over view of the LHAASO detector array at Mt. Haizi (29$^\circ$21'27.6"N , 100$^\circ$08'19.6"E, 4410 m a.s.l.), China. In the middle of the array, the big houses are water Cherenkov detector array (WCDA) of 78,000 m$^2$. In the circular area of 1.3 km$^2$ covered by the web of roads surrounding the WCDA, 5216 scintillator counters and 1188 muon detectors buried under soil humps form the one-square-kilometer array (KM2A). The Panel (B) is the picture of the HAWC array of 300 water tanks at Sierra Negra Volcano (18$^\circ$59'41"N , 97$^\circ$18'30"W, 4100 m a.s.l.), Mexico. The area of the array is $\sim$20,000 m$^2$. Each tank has 4 PMTs at 5 m beneath the water surface. The panel (C) is the picture of AS$\gamma$+MD array, at Yangbajing, (30$^\circ$06'38"N , 90$^\circ$31'50"E, 4300 m a.s.l.). Picture (A) is from LHAASO Collaboration, (B) is from HAWC Collaboration web site https://www.hawc-observatory.org/ and (C) is from AS$\gamma$ Collaboration.}
\label{Fig:LHA-pictures}
\end{figure}

The technique of using particle detector array to detect high energy cosmic ray particle induced cascade process has a long history more than 70 years. Many techniques of particle detection are developed such as the completely covered resistive-plate-chambers used in ARGO-YBJ Experiment \citep{argo}, water Cherenkov detectors in Milagro \citep{MILAGRO-1996}, HAWC \citep{hawc} and LHAASO/WCDA \citep{LHAASO_2010,lhaaso2019} experiments, and more widely used scintillator counters in CASA-MIA \citep{CASA-MIA}, AS$\gamma$ \citep{asgamma} and LHAASO/KM2A \citep{LHAASO_2010,lhaaso2019} experiments. The pictures of the three experiments at high altitudes are shown in Figure \ref{Fig:LHA-pictures}. Timing the secondary particles that register each detector in the array at surface with a precision of $\sim$1\,ns, one can reconstruct the EAS front thus find the arrival direction of the primary particle. The angular resolution strongly depends on how many detectors are registered in an  EAS event. Number of particles recorded in the detectors allows the energy of the primary particle being reconstructed. Fill factor of the active detector area to the total covered area by the array plays an important role to maintain a high resolution. However, the ultimate limit to the angular resolution is set by the intrinsic fluctuations of the arrival time of particles in the shower front.  The high altitudes that the EAS array situated is found crucial as well. Above 4300 m above sea level (a.s.l.), the detector arrays are close enough to the shower maximum, so that the shower fluctuations around 1 PeV are minimized. The shower detection threshold is lowered at higher site as well.  However, to use this technique for $\gamma$-ray detection, the very high level of diffuse CR background is still a major difficulty. Even for the brightest point-like sources, such as the Crab, the photon signal flux is lower than the CR background in the point spread function (PSF) by orders of magnitudes. Until the second decade of 2000's, this essentially sets the limit of sensitivities of the detectors worse than  $\sim$1 Crab Unit (CU) (See the right panel of Figure \ref{Fig:CrabRate}) in $\gamma$-ray detection, even for the mega-scale array like CASA-MIA with a size of 1/4 km$^2$ (not shown in Figure \ref{Fig:CrabRate}) .  
%------------------Fig1

\subsection{CR Background Suppression Techniques and Capabilities}
\begin{figure}[htbp]
%\centering
\subfigure{
\begin{minipage}[b]{0.54\textwidth}
\includegraphics[width=1\textwidth]{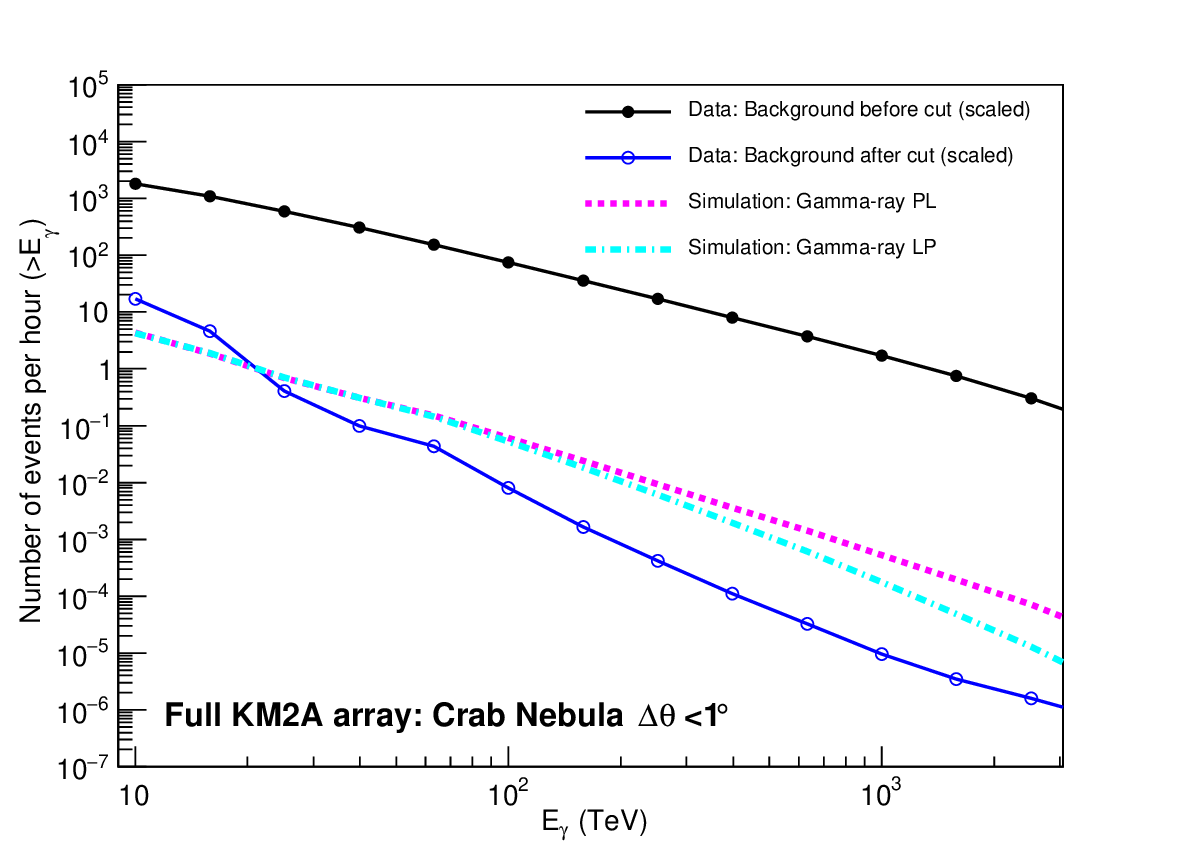}
\end{minipage}}
\subfigure{
\begin{minipage}[b]{0.46\textwidth}
\includegraphics[width=1\textwidth]{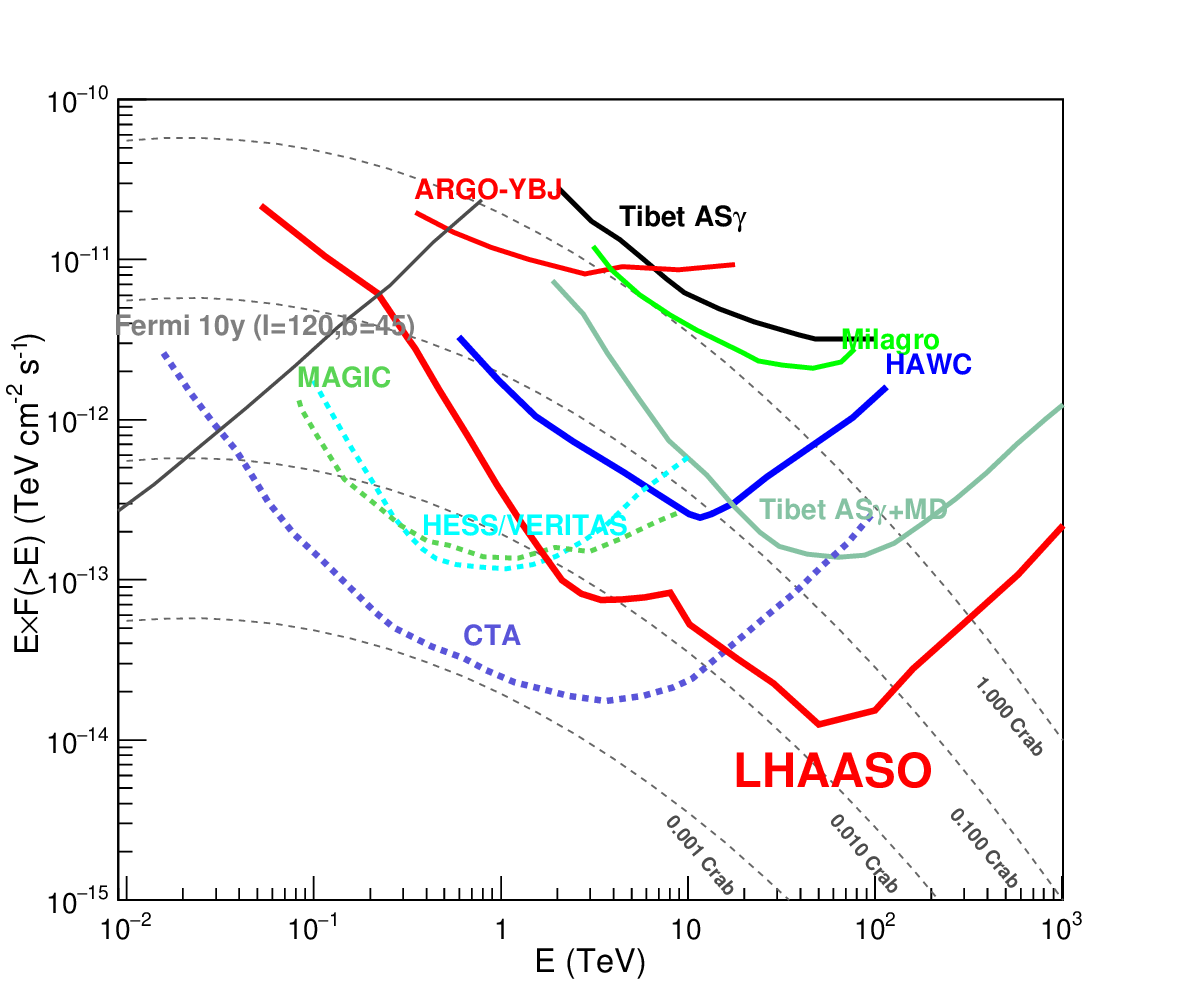}
\end{minipage}
}
\caption{{\bf Left}: The rates of detection of \grays from the Crab and the CR background events above the shower
energy E$_\gamma$ by the LHAASO-KM2A array in a cone of 1$^{\circ}$ centered at the Crab direction. 
The cyan dash-dotted and pink dashed lines represent the integrated  rates   of detected $\gamma$-rays from the Crab,  based on   log-parabola and power-law models fitted to the measured fluxes, respectively. Black filled circles show the integrated rate of cosmic ray events before applying `muon-less cut'. Blue open circles  represent  the integrated rate of remaining cosmic ray events after applying the `muon-less cut' filter.  The figure is from Ref.\cite{LHAASO-Crab} 
{\bf Right}: Sensitivities of VHE and UHE $\gamma$-ray astronomical instruments as functions of $\gamma$-ray energy, E. The Crab Nebula SED in a log-parabola functional form in gray short dashed lines is a global fitting of all the measurements presented in the Fig.3 of \citep{LHAASO-Crab}.  The ground based EAS experiments, Tibet AS$\gamma$, AS$\gamma$+MD \citep{sako09}, ARGO-YBJ \citep{argo_catalog},  Milagro, HAWC \citep{deyoung12} and LHAASO \citep{Cao19}  are represented by colored solid lines. The IACT experiments, CTA \citep{bern13}, VERITAS, HESS\citep{deangelis08}, MAGIC \citep{tescaro14} are represented by colored dotted lines. The 10 year sensitivity of Fermi-LAT \citep{fermi_sensitivity} is represented by the gray solid line. }
\label{Fig:CrabRate}
\end{figure}

In principle, the $\mu$-content of a shower is a clear veto to suppress the CR background. In showers induced by CRs, the multi-particle production generates large number of muons. In contrast, because the very small photo-production cross section in pure electromagnetic cascade induced by a primary photon, only small $\mu$-content is expected. In practice, to realize effective veto by measuring such small $\mu$-content, a quite significant fill factor of muon detectors is required thus it is very difficult. For instance, the fill factor $\sim$1\% of active muon detectors  in CASA-MIA experiment was found not sufficient. An economically affordable solution of $\mu$-content measurement was needed and developed in past two decades. As the second generation of EAS detector arrays,  the AS$\gamma$+MD\citep{ASG_MD-2015} experiment, HAWC as well, combined the two key features of the EAS detection, namely the high altitude ($>$4000 m a.s.l.) and effective $\mu$-content measurement with fill factor of 5\% thus successfully boosted the sensitivity of $\gamma$-ray detection by a factor of 10. Almost at the same time, LHAASO design was approved in 2015 with a combination of 78,000 m$^2$ water Cherenkov detector array (WCDA), which has a typical CR background rejection power of 10$^{-3}$, and an 1 km$^2$ scintillator counter array, in which 1188 muon detectors with 40,000 m$^2$ total active area are uniformly distributed (KM2A). The CR background rejection power reaches to 10$^{-4}$ at 100 TeV and 10$^{-5}$ above 500 TeV, see the left panel of Figure \ref{Fig:CrabRate}. The 15 m spacing between counters in KM2A and 5$\times$5 m$^2$ cells in WCDA enable the angular resolution of 0.2$^\circ$ at 10 TeV by WCDA and above 400 TeV by KM2A. Each detector in the array is equipped with the White-Rabbit protocol based clock distribution system by which the clocks in the detectors are  synchronized with an accuracy of 0.2 ns. The state-of-the-art sensitivity of $\gamma$-ray source detection by LHAASO reaches the level of 0.012 CU as shown in the right panel of Figure \ref{Fig:CrabRate}. With a factor of ten improvements in sensitivity above 50 TeV comparing to the previous generation experiments, LHAASO\citep{LHAASO-ScienceBook} has become the major instrument in the UHE $\gamma$-ray astronomy.

%------------------Fig1
% \begin{figure}[htbp]
% %\centering
% \includegraphics[width=1.0\textwidth]{sensitivity.eps}
% \caption{Sensitivities of VHE and UHE $\gamma$-ray astronomical instruments as functions of $\gamma$-ray energy, E. The Crab Nebula SED in a log-parabola functional form in gray short dashed lines is a global fitting of all the measurements presented in the Fig.3 of \citep{LHAASO-Crab}.  The ground based EAS experiments, Tibet AS$\gamma$, AS$\gamma$+MD \citep{sako09}, ARGO-YBJ \citep{argo_catalog},  Milagro, HAWC \citep{deyoung12} and LHAASO \citep{Cao19}  are represented by colored solid lines. The IACT experiments, CTA \citep{bern13}, VERITAS, HESS\citep{deangelis08}, MAGIC \citep{tescaro14} are represented by colored dotted lines. The 10 year sensitivity of Fermi-LAT \citep{fermi_sensitivity} is represented by the gray solid line. }
% \label{Fig:sensitivity}
% \end{figure}

\subsection{Survey for Sources and Targeted Observation for Deep Investigations}
The field of view of EAS array typically covers 1/6 of the sky  at any moment. The operation duty cycle is typically greater than 95\%. Thus, the EAS arrays are ideal for sky survey for $\gamma$-ray sources particularly for the extended sources. 
%In Fig.\ref{fig:fov} we present the 1 year exposure map of LHAASO experiments. In this regard 
The HAWC collaboration has published the VHE source catalog in the Galactic plane \citep{hawc_3hawc}. LHAASO, which operates in higher energy range, will release the catalog soon. By using a half of designed capacity of LHAASO, 12 Galactic UHE $\gamma$-ray sources are found in 11 months of data taking \citep{lhaaso_nature}. Most of the sources  are found extended, so  multiple individual objects could be potentially associated with each of the UHE \gray sources.  A better angular resolution than the EAS array is needed for further investigation the origin of the UHE photons. The $r_{\rm 68}$ of the point spread function (PSF), which is defined as the radius inside which 68\%  of the photons from the point source are contained, is about $0.2^{\circ}$ for HAWC above $10~\rm TeV$ and LHAASO KM2A above $100~\rm TeV$ \citep{lhaaso_km2a}. 
%PSF is also highly energy dependent and at lower energy $r_{\rm 68}$ would be much larger. Most of the discovered UHE and VHE $\gamma$-ray sources locate inside the Galactic plane, which is a extremely crowded area. Thus the $0.2^{\circ}$ PSF is not sufficient to disentangle those very near sources. 
The  $r_{\rm 68}$ of imaging air Cherenkov Telescope arrays (IACTs) is typically 2 arcminutes, which is 5 times better than those of EAS arrays. Thus IACTs are ideal instruments to complement with the EAS arrays for targeted observations. However, due to the small effective acceptance of the existing IACTs, 
there is yet no source firmly detected by IACTs in UHE domain. The next generation IACTs such as CTA  and particularly the newly proposed ASTRI\citep{astri-2022} and LACT, will be equipped with large number of telescopes to enhance the collection area up to $\sim$$10^6~\rm m^2$ and have good synergy with current EAS arrays, and will perform targeted observations in depth towards the UHE sources. The other aspect of the UHE $\gamma$-ray astronomic observation is only the northern sky is covered by the surveying instruments. Newly proposed Southern Wide-FoV Gamma-ray Observatory (SWGO)\citep{SWGO-2022} is an EAS detection instrument located in a high altitude site in southern hemisphere having similar or even better sensitivity than LHAASO. Many UHE sources are expected to be discovered in the inner part  of our galaxy including GC.

\section{DISCOVERY OF PEVATRONS}
One of the most important topics of $\gamma$-ray astronomy is to search for PeVatrons. Progresses have been made in past years. The first hint was provided by 
HESS which measured a hard SED up to 20 TeV \citep{hessgc}. Lately, more direct clues were from AS$\gamma$ and HAWC with handful UHE \grays collected and a photon at $\sim$0.4 PeV\citep{ASgamma2019,HAWC_56TeV} recorded by the former. The concrete evidences about PeVatrons were achieved by LHAASO with 534 UHE \gray photons detected, among them 1.4 PeV is the most energetic one \citep{lhaaso_nature} by 2021. Both the highest energy and number of UHE photons will have been renewed by the time when the paper is published. 
In this section, we review the search for PeVatrons, and also discuss various types of possible  astrophysical candidates of the PeVatrons. Three specific promising PeVatron candidates will be further discussed in two subsequent sections.

\subsection{The First Hint from the Galactic Center}
As mentioned in Sec.2.2, the method used before LHAASO to hunt PeVatron is to search for the hard $\gamma$-ray spectrum  above $\sim 100~\rm TeV$ without significant softening. Before 2021, the strongest hint came from the H.E.S.S observations on GC region \citep{hessgc,veritas_gc}. The VHE emission in GC can be decomposed into three components, one bright central point source, the point source associated with SNR G0.9+0.1 and the diffuse emission associated with the gas distributions \citep{hessgc06}. The spectrum of the central point source has a significant cutoff at several TeV. However, the spectrum of the diffuse emission shows a hard spectrum (index of about 2.3) and  no hint of cutoff up to more than $20~\rm TeV$ \citep{hessgc,veritas_gc}. The diffuse $\gamma$-ray spectrum indicates that the 90\% lower limit of the cutoff in  parent CR protons is $0.6~\rm PeV$.    Furthermore,  H.E.S.S collaboration also derived the CR radial distribution with respect to GC by using the $\gamma$-ray flux measurement and the gas distribution based on molecular line emissions. The derived CR spatial distribution is consistent with an $1/r$ profile, which is expected assuming the CRs are injected continuously from the central region with the size of tens of pc at the center of our Galaxy. The possible accelerator may be the supermassive black hole (SMBH) Sagittarius A* itself \citep{hessgc}, or the young massive star clusters, such as Arches, Nuclear cluster and Quintuplet which all locates in the central region. 

Although H.E.S.S results found strong hint on the possible PeVatron in GC region, we note that the  90\% lower limit of the cutoff in  parent CR protons is $0.6~\rm PeV$ while 95\% lower limit is of only $0.4~\rm PeV$. Indeed, the cutoff energy $E_c$ in $\gamma$-ray spectrum reflects the cutoff in the parent proton spectrum of about $10-20$ $E_c$ \citep{kelner06}.  And MAGIC observation \citep{magic_gc} in the same regions found hints for spectral cutoff, which is in contradiction with the measurements of H.E.S.S and VERITAS.  More solid identification of PeVatrons requires the accurate spectral measurement above $100~\rm TeV$, that is, the UHE domain in which  IACTs does not have sufficient sensitivity.  

\subsection{Discovery of the First Group of PeVatrons with the Flux $\sim$1 CU }
Before 2021, several Galactic sources were observed with $\gamma$-ray energies slightly higher than 0.1 PeV by the AS$\gamma$ \citep{ASgamma2019}, HAWC \citep{HAWC_56TeV} and MAGIC \citep{magiccrab2020} experiments. These observations provide more direct hints of existence of Galactic PeVatrons. However, unbiased identification and in-depth investigation of PeVatrons require detection of steady \gray fluxes with energies well above 0.1 PeV for hadron PeVatrons. 
%, the typical energy of the secondary \gray is about 1/10 of the parent proton energy that requiring \gray energy above 0.1 PeV.
Alternatively, a stable $\gamma$-ray flux above 0.4 PeV must be detected for a lepton Pevatrons, if it existed, where the IC scattering of a parent electron generates $\sim 0.4(E_e/\rm PeV)^{1.3}$ PeV photons with energy of $E_e$. 
%by the parent electron energy, that requiring . Therefore, observation of $>$0.4 PeV \gray could provide model independent evidence for the identification PeVatrons. 

\begin{figure}[htbp]
%\centering
\includegraphics[width=0.9\textwidth,height=4.5 cm]{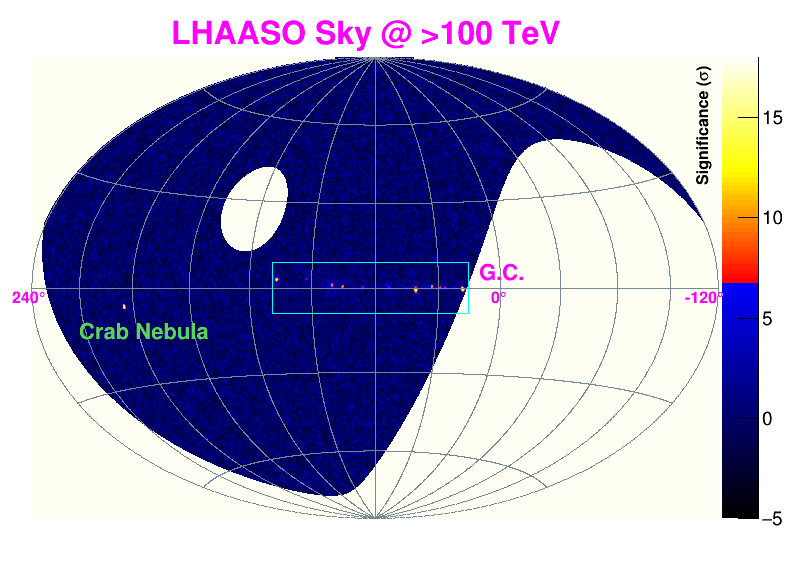}
\includegraphics[width=1.0\textwidth,height=3.0 cm]{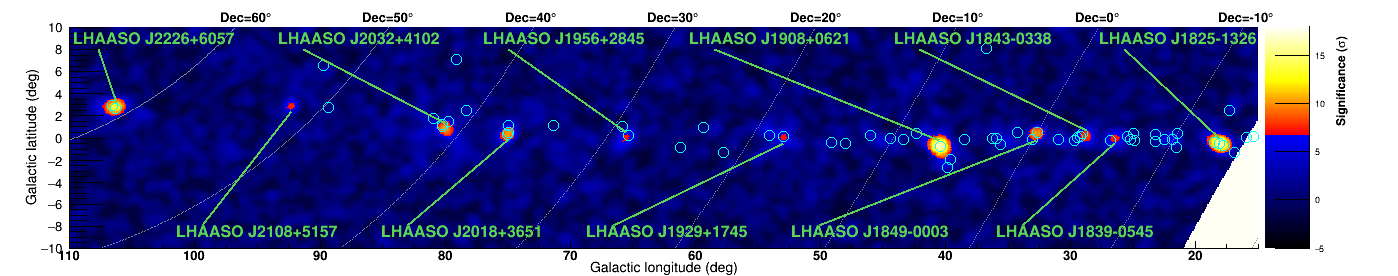}
\caption{  LHAASO sky map at energies above 100 TeV. The circles indicate the positions of known VHE \gray sources. The figure is from Ref.\cite{lhaaso_nature}.} 
\label{Fig:Sky}
\end{figure}

The LHAASO collaboration reported the detection of more than 500 photons at energies above 100 TeV that form 12 clear clusters in sky with statistical significance $>$7$\sigma$ for each of them, thus revealed ultrahigh-energy \gray sources\citep{lhaaso_nature} in 2021.  This marked the discovery of the first group of PeVatrons. The most energetic $\gamma$-ray was found at 1.4 PeV from the Cygnus region.  As shown in the significance map in Figure \ref{Fig:Sky}, the 12 sources lined up with  good coincidence with the Galactic plane. Most of those UHE $\gamma$-ray sources are associated with known VHE sources. This hints that most likely the Milky Way is full of PeVatrons. 
%In further deeper survey with LHAASO, many more PeVatrons are expect to be revealed in near future. 

Within the 12 UHE sources, eight sources were found emitting \grays more energetic than  0.4 PeV.  Several potential counterparts are found in their proximity, including PWNe, SNRs and star-forming regions, however, except for the Crab Nebula, the firm identifications of production sites have not been established, yet \citep{lhaaso_nature}.  Further investigations in depth, particularly multi wavelength analyses, will help identifying the relevant candidates accounting for the PeVatrons. 

LHAASO also measured SEDs of three most luminous sources, i.e. LHAASO J1825-1326, LHAASO J1908+0621 and LHAASO J2226+6057. Despite of steep spectra of these sources, no clear cutoff features are found below 500 TeV. For all sources, the absorption due to ISRF and CMB is found small, even for the photons at the highest energies. The SED of $\gamma$-rays, almost directly represent the parent particle energy distributions in the PeVatrons, so that they are crucial for revealing the corresponding particle acceleration mechanism.  Further phenomenological studies have been  following up in literature  to localize and identify the PeVatrons. Here is a brief summary of them.

\subsection{Possible Astrophysical counterparts of PeVatrons}
\subsubsection{Pulsar Wind Nebulae}
PWNe are powered by energetic pulsars, which are composed of electrons and positrons driven from the magnetosphere. These particles form a cold ultrarelativistic wind and are further accelerated at the termination shock generated when the pulsar wind encounters the ambient medium \citep{gaensler06}. PWNe have been recognized as one type of the most efficient electron factories in the Galaxy. A large fraction of identified Galactic VHE sources are PWNe. It is widely believed that the high-energy \gray emission of PWNe mainly comes from IC scattering of the high-energy electrons/positrons on ambient low-energy photons, such as the interstellar infrared radiation field and the CMB. However, hadronic processes are also suggested responsible to the $\gamma$-ray emissions, particularly at high energy ends of SEDs (see detailed discussion in Section~\ref{sec:crab_had}).

Among the twelve UHE sources detected by LHAASO, the only firmly identified is the Crab Nebula well known as a PWN\footnote{Crab Nebula, as well as some other PWNe, is sometimes also called an SNR because it is formed after the supernova explosion. We here refer to it as a PWN based on the physical origin of its radiation, which is produced by electrons/positrons blown from the pulsar}. In a deep investigation, LHAASO reported the SED of the  Crab Nebula  extending to 1.1 PeV following a simple log-parabola functional form  with index of -3.12$\pm$0.03 around 1 PeV \citep{LHAASO-Crab}.  A detailed review on the Crab Nebula can be found in next section. 
%Therefore, The PWNe can been firmly identified as the extreme electron PeVatron. 

In the vicinity of every single LHAASO-detected UHE sources, except for LHAASO J2108+5157, there are at least one energetic pulsars with spin-down power above 10$^{35}$ erg/s (more detailed discussion can be found in Ref.\cite{wilhelmi22}, according to which the maximum electron energy derived from the spin-down power of the pulsars ranging from 1\,PeV to 10\,PeV do not contradict to the observation of LHAASO, except for the one in the Cygnus region, i.e., LHAASO J2032+4102. Such a strong correlation indicates that PWNe are very likely responsible to the UHE emission of those PeVatrons, thus there might be potential candidates of the PeV accelerators among them. 

Recent observations on the HAWC J1826-128 \citep{burgess22}, spatially coincident with LHAASO J1825-1326, and MGRO J1908+06 \citep{hawc_1908}  coincident with LHAASO J1908+0621, suggest PWNe to be responsible to the VHE emission. The emission from LHAASO J2226+6057 \citep{desarkar22, liang22}  and eHWC J2019+368 \citep{fang20}, coincident with LHAASO J2018+3651, were also explored theoretically in the scenario of PWN. However, all the investigations are all yet to be conclusive, with the competitive radiation mechanism not being ruled out. Further measurements are still highly desired with better statistics.   

\subsubsection{Young Massive Star Clusters}
Young massive stars, which generate strong star winds, in a dense cluster may form multiple shocks with a potential to accelerate CR protons to very high energies above 1 PeV, as suggested by authors \cite{aharonian19}. The YMCs are recognized as the major factories of Galactic CRs with many evidences in VHE domain. The positional coincidence of LHAASO J2032+4102 with the YMC Cygnus OB2 provides further evidence for the YMC to be a hadronic PeVatron. A recent report based on the HAWC observation also attributes the emission at energies from  1 TeV to 100 TeV to the enclosed star-forming region Cygnus OB2 \citep{hawc_cygnus}. Another possible evidence may be from the positional coincidence of LHAASO J1849-0003 with W43. However, further morphological analysis in depth is deserved to clarify the association based on the future data collection.

\subsubsection{Supernova remnants}
SNRs, the spherical shock waves expanding in the ISM after the explosion of massive stars, have been proposed as the most promising sources of Galactic CRs for long time. 

%The detection of a characteristic pion-decay feature at sub-GeV energy provides direct evidence that protons are accelerated in SNRs \citep{fermi_pion}. This, however, tells nothing about the acceleration capability limit of SNR at the highest energy. 
The detection $\gamma$-rays above 100 TeV from SNRs would give clues on the acceleration capability limit of SNRs at the highest energy. Evidences for a SNR to be a PeVatron are very crucial for understanding the origin of CRs in the knee region. In the VHE domain, all the SEDs of young SNRs appear to be quite steep or have breaks at energies below 10 TeV. This has raised doubts about the ability of SNRs to operate as PeVatrons \citep{aharonian19}. More details about SNRs as candidates of PeVatrons can be found in a recent review article \citep{cristofari21}.

Among the LHAASO UHE sources, 6 out of 12 are found having a SNR in the vicinity of them. However, most of the spatial coincidences are competing with energetic pulsars, and usually the PWN scenarios are more preferred. The most favorable PeVatron candidate would be LHAASO J2226+6057  which is likely associated with SNR G106.3+2.7. Detailed discussion about the recent progresses concerning the SNR G106.3+2.7 region can be found in Section 6.2. It is worth noting that further morphological analysis for this source at UHE band in near future might provide crucial information for the identification of the PeVatron. %Recent AS$\gamma$ observation at energies $>$10 TeV indicates that the emission region might be correlated with the surrounding Molecular Clouds (MC) rather than with the pulsar PSR J2229+6114 \citep{ASg_2228}. However, this result has some difference with the observation of HAWC \citep{hawc_g106} and LHAASO \citep{lhaaso_nature} at higher energies. In the VHE domain, the newly MAGIC observation shows that the emission was divided into two morphological regions, i.e. the head and the tail regions \citep{magic_g106}. The head region is likely associated with Boomerang PWN, and the tail region is likely associated with SNR G106.3+2.7 and the MCs in the region. Similar observation is also completed in X-ray band \citep{Ge21}, which had been taken as the evidence to support the SNR G106.3+2.7 to be a powerful proton accelerator up to 1 PeV. Further morphological analysis for this source at UHE band shortly will provide crucial information for the identification of the PeVatron, either the SNR or the PWN. 

\subsubsection{Micro-quasars}
A micro-quasar consists of a binary system of a compact object (either a black hole or a neutron star) accreting matter from a companion star. Such a miniature system can display  some of the properties of quasars with relativistic jets. Observation of $\gamma$-rays from jets could provide valuable probes of the particle acceleration mechanisms in the jets. A few of those objects have been detected with $\gamma$-ray emission at high energy band by AGILE and Fermi-LAT \citep[see][and reference therein]{bodaghee13}, e.g. Cygnus X-1 and Cygnus X-3. The $\gamma$-ray with the highest energy around 20 TeV is detected by HAWC from jets of the micro-quasar SS 433 \citep{hawc_ss433}, therefore, micro-quasars could be PeVatron candidates. However, SS 433 is in the vicinity of the very extended source LHAASO J1908+0621, which is so bright that some contamination might be expected and needs to be carefully disentangled in the analysis of SS 433. Micro-quasar Cygnus X-3 has the similar problem because it is almost in the heart of a very complex extended source in the Cygnus region. Even worse,  there is a very bright source LHAASO J2032+4102 very nearby as well. Not only the multi wavelength morphological analyses of those sources are very necessary, but also temporal structure of the emission, particularly in UHE regime, would play a crucial  role in further detailed investigation. Such a combined analysis may shed light on the identification of micro-quasars as candidates of PeVatrons. 

\section{The Crab Nebula: an Extreme Electron Accelerator and a Potential Super-Pevatron}
Here, we discuss the Crab Nebula as the first well-studied PeVatron. We will analyze the potential for astrophysics discovery and impact on $\gamma$-ray astronomy through targeted observations of this special lepton-PeVatron.  

On July 4th, 1054, Chinese astronomers recorded the supernova that evolved into today's Crab Nebula, the best observed high-energy astrophysical object. The ejecta forms the remnant with a size of  $\sim$11 light years (ly). The central pulsar with the spin period of 33 millisecond powers strong wind of electron-positron pairs with the spin-down luminosity 4.6$\times 10^{38}$ erg/s. This forms clear torus structure of termination shock fronts at radius of 0.59 ly (inner) and 1.49 ly (outer). They are very bright in X-ray band, not only the rings, but also clear structure of knots and a pair of jets indicate regions of strong radiation. All those together with diffuse radiative region surrounding the pulsar forms a nebula of $\sim $3 ly \citep{Ng04}.

Relevant feature of the Crab Nebula is its radiation covers nearly the whole electromagnetic wavelength range  from radio to the highest $\gamma$-rays at $\sim$1 PeV. The one-zone leptonic model roughly describes the main feature of the SED over 22 orders of magnitudes by assuming a bulk of electrons, confined by an average magnetic field of $\sim$100 $\mu$G, emitting photons up to 1 GeV via the synchrotron radiation and generating higher energy $\gamma$-rays through the IC scattering process, as shown in the left panel of Figure \ref{Fig:Crab-SED}. However, statistic test on the agreement between the model and data shows that there are systematic deviations in almost all specific bands, e.g. the whole energy range covered by Fermi-LAT and particularly in the UHE band. The strong systematic deviation indicates possibility of existence of new components.

% Sec. 5.1
\subsection{Facts and Challenges}
The first evidence of UHE photon emission from the Crab Nebula was from HAWC \citep{HAWC_crab} , AS$\gamma$ \citep{ASgamma2019}  and MAGIC \citep{magiccrab2020} experiments using various techniques. Handful UHE photons are collected.

The systematic observation has been done by LHAASO \cite{LHAASO-Crab} which collected 89 UHE photons from the Crab, including two photons at 0.9 and 1.1 PeV. Together with the photons measured by LHAASO WCDA and KM2A at lower energies,  a  log-parabola   spectrum   over 3 decades of energy clearly reveals the radiation feature of the Crab Nebula in the UHE domain.
The PeV photons are measured with negligible  probabilities of CR background contamination.

Assuming the PeV photons being generated by electrons, a couple of characteristics are summarized as follows. A) the parent electron must have the energy of 2.3 PeV. B) The smallest region for those electrons to be confined in the magnetic field of 110 $\mu$G is 0.08 ly. C) Acceleration rate could be as high as 16\%, which is a factor of 1000 higher than diffuse shock acceleration in SNRs.

The SED analysis demonstrates that the one-zone model seems to be too simple. The systematic deviations at a level more than 10$\sigma$ are found in various bands from radio to UHE $\gamma$-rays. From  1 GeV to 1 PeV, the $\gamma$-ray spectrum could be much better fitted by introducing the second component of sources, either electrons or protons \citep[e.g.][]{Nie-2022}. It was found that the systematic deviations are nearly completely removed in the SED fitting over the whole $\gamma$-ray band by simply introducing a proton component for the spectrum above 300 TeV, as shown in the left panel of Figure \ref{Fig:Crab-SED}. There is more detailed discussion below.

Many progresses have been made in modeling of the Crab Nebula based on the magnetohydrodynamic (MHD) calculation using particle-in-cell (PIC)  technique by various research groups \citep[]{Lyubarsky-2007, Amato-2021-Crab,Du-2021}. Many details of plasma evolution, terrain of the shock front and particle acceleration have been revealed in 1D and 2D simulations. The capacities of modern computing facilities allowed the exploration even in 3D domain. This helped to understand how particles are accelerated by tracing them through the tabulating plasma. Some fundamental questions, such as the extremely large acceleration rate observed in the Crab Nebula, are still open. It is still difficult to understand the acceleration of electrons up to the level of 1 PeV \citep{French-2021}. The lower energy radiation, in the bands of radio and GeV $\gamma$-rays, would not be explained well in the same theoretic frame if the initial conditions of the simulations were pushed too high for generating the extremely high energy photons.

% Sec. 5.2
\subsection{A Super-PeVatron of protons?}
\label{sec:crab_had}
\begin{figure}[htbp]
\subfigure[]{
\begin{minipage}[b]{0.48\textwidth}
\includegraphics[width=1.0\textwidth]{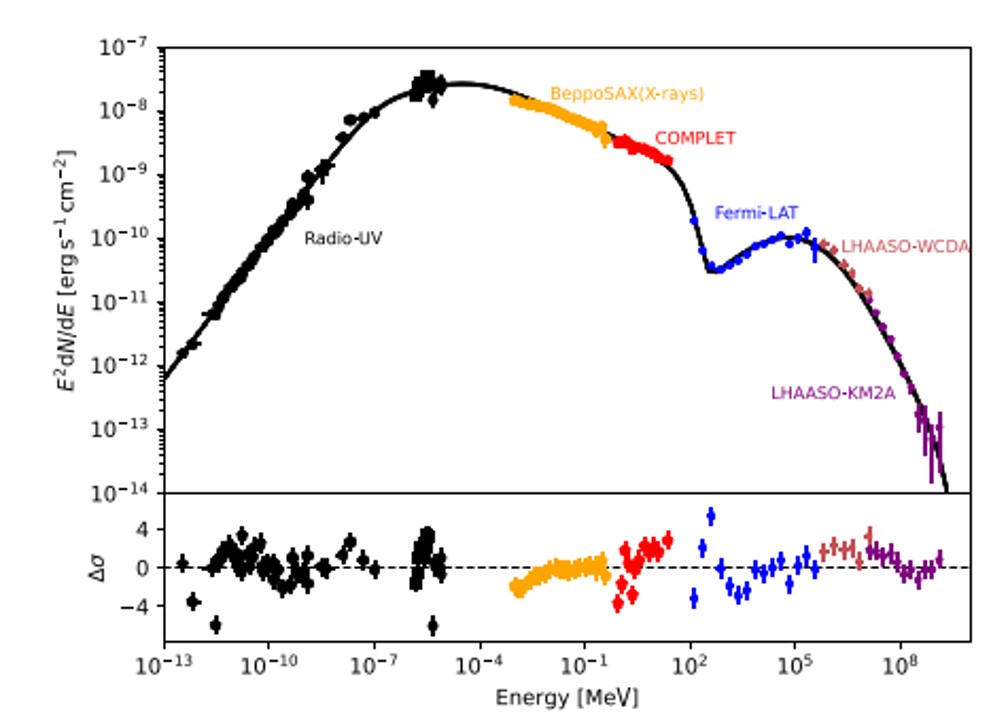}
\end{minipage}
}
\subfigure[]{
\begin{minipage}[b]{0.48\textwidth}
\includegraphics[width=1.0\textwidth,height=4.3
cm]{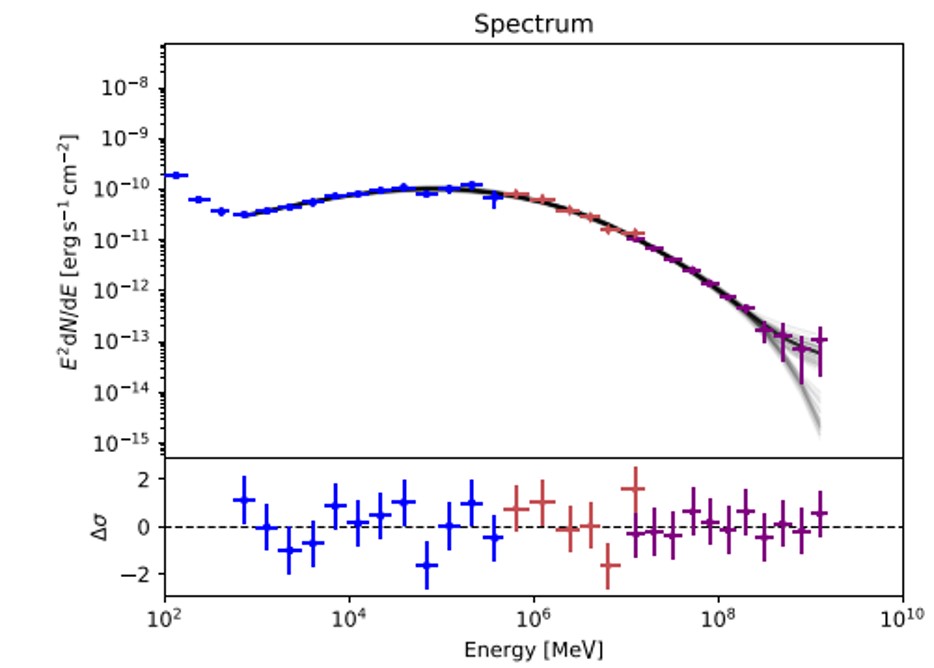}
\end{minipage}
}
\caption{Left: The SED of the Crab Nebula fitted with a simple one-zone electron model, where the names of experiments that contributed the data in different bands in specific color are marked. The lower panel shows the deviation of the model from the data. 
Right: The combined SED of the Crab Nebula with Fermi-LAT in bulue and LHAASO/WCDA in red and LHAASO/KM2A in purple in the upper panel. The SED is well fitted above 1 GeV using a hybrid model with the `standard' one-zone lepton component plus a high energy proton component, indicated by the shaded gray curves. In the lower panel, the deviation of the model from the data is plotted in the unit of standard deviation $\sigma$. A rather good agreement is clearly shown in this plot. Both panels are taken from Ref.\cite{Nie-2022}.
%Right: The illustration of the hybrid model by using the $\gamma$-ray flux contributions of all sub-components in the one-zone leptonic model in dashed and dotted curves, plus a very energetic proton component with a cut-off at 30 PeV in blue solid curve. All sub-components are explained in legend and more detailed information can be found \citep{Nie-2022} from wich the panels are referenced. Data from BeppoSAX in yellow and COMPTEL in red in X-ray band are also plotted. However, due to unclear possibility of flaring of the Crab Nebula, the COMPTEL and low energy part of Fermi-LAT were not taken into consideration in the fitting procedure. 
}
\label{Fig:Crab-SED}
\end{figure}

Although it is generally believed that most of the rotational energy of pulsars reduced during its spin-down is converted into the energy of electron/positron pairs and magnetic fields in the PWNe, protons can be also loaded in the pulsar wind and accelerated to high energies. In fact, there have been some studies discussing proton acceleration and pionic $\gamma$-ray signature in PWNe, especially in Crab Nebula since decades ago \citep[e.g.,][]{Nie-2022, Cheng90, Atoyan96, Amato03, horns06, Zhangl09, DiPalma17, ZhangX20, Peng2022, liang22}. The possible structure of SED of Crab Nebula at the highest energy, yet to be statistically significantly confirmed in LHAASO's observation for years, may be accounted for by the hybrid model of a `standard' one-zone leptonic origin plus a very energetic proton component with a cut-off at the energy $\gtrsim 10\,$PeV, as %illustrated 
shown in the right panel of Figure \ref{Fig:Crab-SED}. Using the last two data points as upper limits to constrain the flux of the hadronic component, 
%Liu \& Wang 
it is found that protons could take  $10\%-50\%$ of the released rotational energy as the maximum considering protons having to escape from the nebula \cite{Liu21_crab}. If each pulsar in our Galaxy converted a comparable fraction of rotational energies to protons, pulsars might potentially explain the CR flux in the range of $10-100$\,PeV.

\section{The Galactic CR factories}  %Molecular Clouds illuminated by a Super-PeVatron }
Supernova remnants are believed to be the most prominent CR accelerators in our Galaxy, AGILE \citep{agilesnr} and Fermi-LAT \citep{fermi_pion}   have revealed  distinct low energy breaks in \gray spectrum in mid-aged SNRs (with an age of about 10000 years). Such spectral features are regarded as a decisive proof that SNRs do accelerate CRs. On the other hand, GeV and TeV \gray observations also found the young massive star clusters (YMCs) such as Cygnus OB2 \citep{fermi_cygnus,hawc_cygnus, aharonian19}, Westerlund 1 \citep{hess_w1}, Westerlund 2\citep{yang18,hess_w2} are also potential CR accelerators. Furthermore, the spectral feature reveals that these YMCs can even be promising PeVatron candidates \citep{aharonian19}. In this section we summarize the current status and discuss the  prospect of the UHE observations of those objects.

% Example of lists
\subsection{Cygnus Region: An Ideal Astrophysics Lab}
Young massive clusters have long been regarded as candidates of CR accelerators \citep{cesarsky83}. Thanks to the advance of \gray instruments, nearly a dozen of young massive clusters are detected in \grays\citep{hess_w2,hess_w1,fermi_cygnus,yang16, yang18,sun20a,yang20w43}.  Amongst them, Cygnus region, due to its proximity and high luminosity, is the best studied young massive cluster system.  
Cygnus region is one of the most intensive and nearby (at a distance $\approx 1.4$~kpc) star-forming regions in our Galaxy. It harbors several Wolf-Rayet and hundreds of O-type stars grouped in powerful OB associations. It also contains huge  HI and molecular gas complexes with the total mass of more than  $10^{6} M_\odot$. 
TeV \grays have already been detected by HEGRA\citep{hegra_cygnus}, which is the first unidentified source in \gray band.  
Fermi-LAT  detected the high energy (GeV) $\gamma$-rays  from the direction of the most massive star association Cygnus OB2. The source  with $\sim 2^\circ$ extension (dubbed `Cygnus Cocoon' \citep{fermi_cygnus}) later has been also detected in the TeV band\citep{Cyg_argo,hawc_cygnus}. The SED of the cocoon shows a hard spectrum (with an index of about 2.2) below 1 TeV and a gradual softening in the TeV band. Such a spectral feature can be explained by either propagation effects assuming a recent injection within 0.1\,Myr or a cutoff in the injected CR spectra \citep{hawc_cygnus}. 
Furthermore, the CR spatial distribution derived from both the GeV and TeV \gray surface brightness \citep{aharonian19,hawc_cygnus}  and gas distributions in Cygnus cocoon obeys  a $1/r$ profile, which is  consistent with the continuous injection of CRs, where $r$ is the distance to Cygnus OB2,  the most probable CR accelerator in this region. Cygnus OB2 is one of the most powerful young massive cluster in our Galaxy, which is consist of more than 50 O-type stars and the total wind mechanical power is estimated as $10^{39}~\rm erg/s$. Assuming a reasonable acceleration rate (10\%), combining the size of Cygnus cocoon and the derived CR energy density profile, the diffusion coefficient inside cocoon is estimated to be at least a factor of 100 smaller than the fiducial value in the Galactic plane \citep{aharonian19}.   

Above $100~\rm TeV$, the HAWC observations didn't show significant cutoff, thus we expect such extended structure should also be detected at UHE band. Recently, AS$\gamma$ reported UHE \gray emission from the Cygnus region \citep{asgamma_cygnus}, with only two compact sources are detected. However, it should be noted that in the diffuse emission detected by AS$\gamma$ at least 4 UHE photons are in the vicinity of Cygnus region \citep{asgamma_diffuse}.  Remarkably the highest-energy photon detected so far, a photon of 1.42$\pm$0.13 PeV (which is a clear identification of photon initiated shower with a probability of 0.028\% to be induced by a background CR \citep{lhaaso_nature}), is found from LHAASO J2032+4102, an extended UHE source in the direction of Cygnus region \citep{lhaaso_nature}. This makes it the most promising PeVatron candidate and provides strong indication of a super-PeVatron which produces CRs above 10\,PeV in our Galaxy. %A deeper observations by LHAASO would be crucial to understand the \gray emission in this regions. 
 
 An obvious question  is  whether the measured size of the Cygnus cocoon is a physical boundary or just caused by the limited sensitivity of instruments. Indeed, in the continuous injection scenario the $1/r$ CR profile predict dimmer surface brightness at large $r$. It is possible that more sensitive instruments would reveal even more extended structures than the cocoon. In this regard, with the unprecedented sensitivity above 100 TeV and large FOV, LHAASO will provide unambiguous information on the \gray spectral and spatial properties in Cygnus region, and shed light on the origin of CRs and identification of the Galactic PeVatron.  

It is worth noting that the Cygnus region is a complex region crowded by SNR Cygni, $\gamma$-ray binary Cygnus X-3 and PSR J2032+4127, except for the YMC Cygnus OB2. Many new observations in VHE and UHE bands are still unclear, e.g. the Carpet-2 experiment team \citep{dzhappuev21}  recently reported the detection of a 3.1$\sigma$ excess of $\gamma$-ray flux at energies $>$300\,TeV  might be associated with a 150\,TeV neutrino event detected by IceCube \citep{icecube20} and is likely consistent with a flare with the duration of a few months.  Therefore, adequate photon statistics provided by LHAASO for spectrometric and morphological studies of this region is desperately expected to address many open questions related to the PeVatron in this region.  

\begin{figure}
%    \centering
    \subfigure{
    \begin{minipage}[b]{0.48\textwidth}
    \includegraphics[width=1\textwidth]{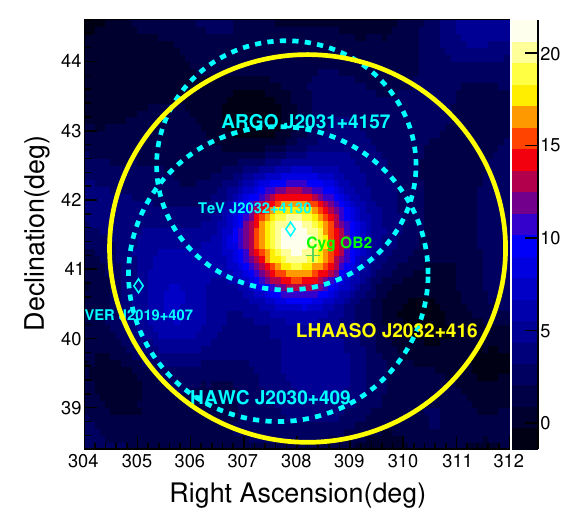}
    \end{minipage}
    }
    \subfigure{
    \begin{minipage}[b]{0.48\textwidth}
    \includegraphics[width=1\textwidth]{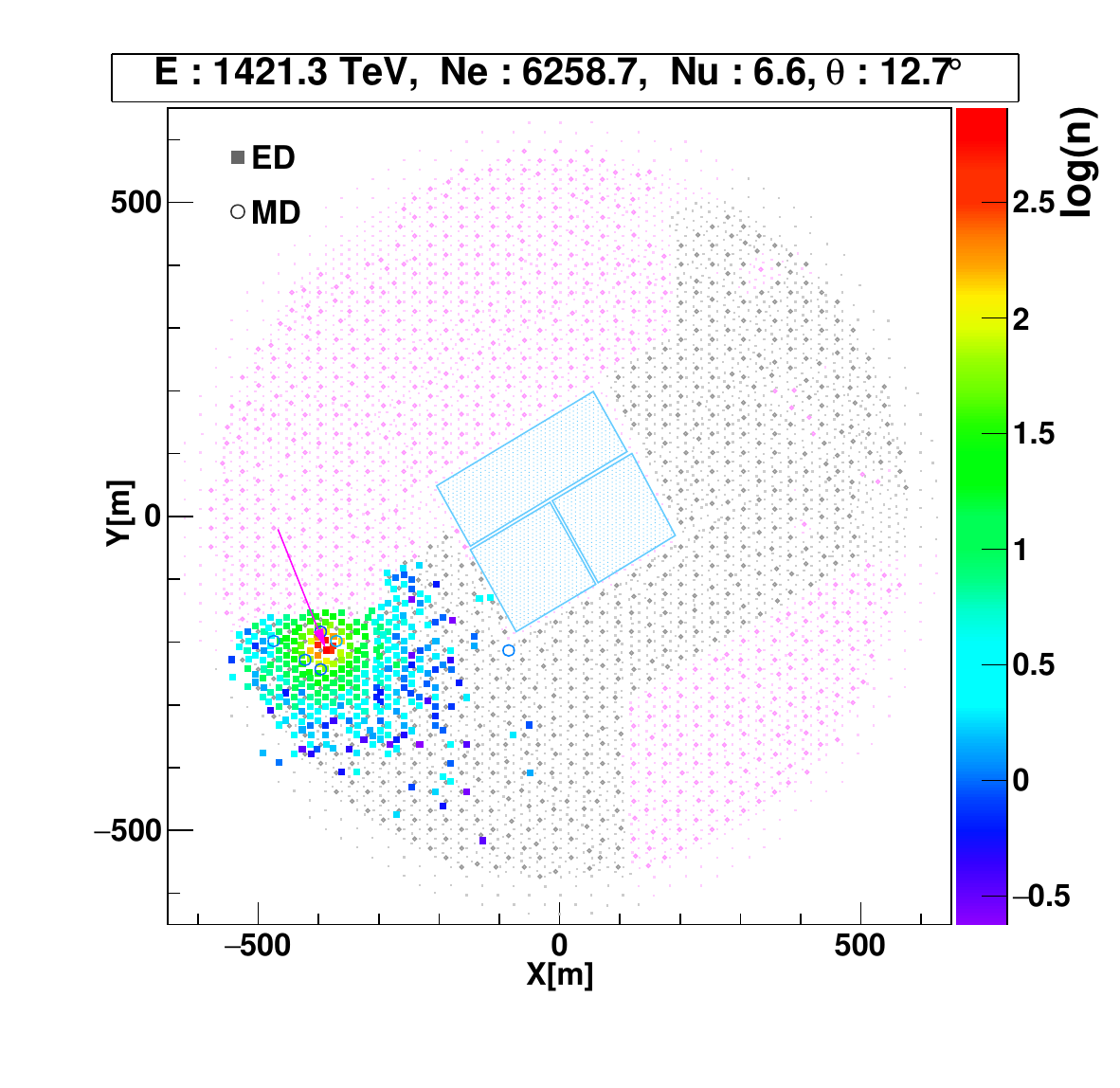}
    \end{minipage}
    }
    \caption{{\bf Left}: The significance map in Cygnus region above 25\,TeV observed by LHAASO. The blue diamonds marks TeV sources TeV J2032+4130 and VER J2019+407. The two blue dashed circle marks two very extended sources ARGO J2031+4157 and HAWC J2030+409. The yellow circle marks the source LHAASO J2032+416. {\bf Right}: The charge distribution for the highest gamma-ray event (1.4\,PeV) detected by LHAASO from the Cygnus region. Both two panels are taken from Ref.\cite{LiC_2021ICRC}}
    \label{fig:cygnus}
\end{figure}

%\begin{enumerate}
%\item The Central region with very strong and energetic $\gamma$-ray radiation
%\item A quite extended bright $\gamma$-ray bubble well associated with gas distribution
%\item A known SNR, the $\gamma$-Cygni
%\end{enumerate}

%\subsection{Molecular and HI Clouds: VHE/UHE $\gamma$-ray Sources}

% Heading 4
%\subsubsection{Molecular/HI cloud observation}

%\subsubsection{Association of the VHE/UHE $\gamma$-ray fluxes  with clouds}

%\subsubsection{Existence of the central factory of cosmic rays}

%\subsection{Micro-quasar:  X-3}
%\subsection{Very Mystery Central Region with $\gamma$-ray Emission above 1 PeV}

%\section{PWN as UHE $\gamma$-ray Sources}
%\subsection{The Crab Nebula and Boomerang}
%\subsection{SNR and Molecular Clouds}

\subsection{SNR G106.3+2.7 as PeVatron Candidate}

G106.3+2.7 is a radio source identified as a SNR \citep{Pineault00}. It presents a quite complex morphology, which can be generally divided into a compact `head' in the northeast part of the source and an elongated `tail' extending towards the southwest. An energetic pulsar, namely, PSR~J2229+6114, is located in the northern part of the head region, surrounded by a boomerang-shaped radio nebula. The latter is named after the morphology as the Boomerang Nebula and believed to be powered by PSR~J2229+6114, which is of a characteristic age of 10\,kyr with the spin-down luminosity of $2.2\times 10^{37}\rm erg/s$. Although yet concretely confirmed, the Boomerang Nebula and SNR G106.3+2.7 are usually considered born from the same supernova explosion. The distance of the system is suggested to be 0.8\,kpc \citep{Kothes01}, based on the apparent spatial correspondence between the radio contour and the distribution of the HI emission around the head region. It appears that the SNR head including Boomerang Nebula is interacting with ambient atomic hydrogen gas while the SNR tail is expanding into a cavity. However, a much farther distance of 3\,kpc of the source is proposed  \cite{Halpern01} based on the hydrogen column density obtained from the X-ray spectral fitting of PSR~J2229+6114.  This could imply a different scenario of the radiation mechanism.

\begin{figure}
    \centering
    \includegraphics[width=1\textwidth]{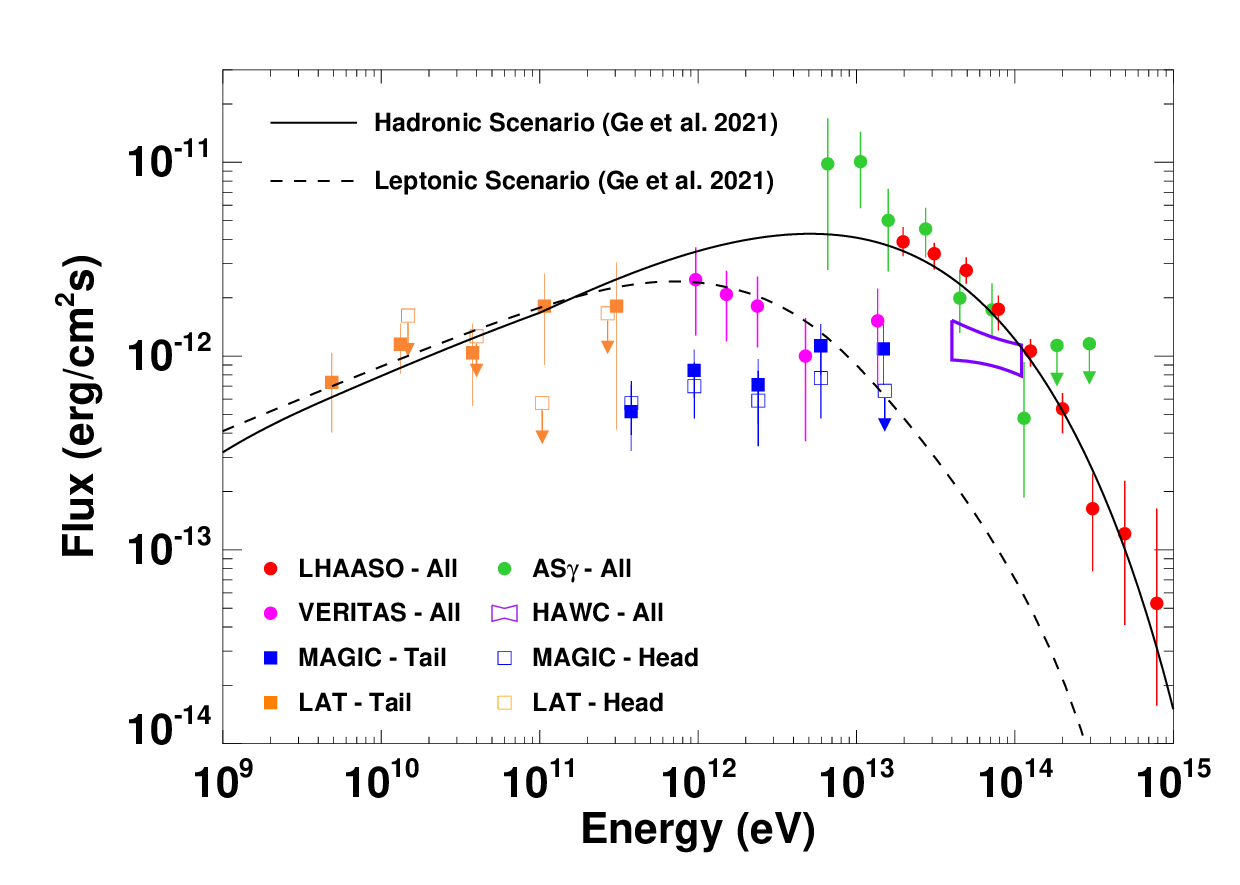}
    \caption{GeV -- PeV \gray SED of G106+2.7 region measured by various instruments. LHAASO's data is taken from Ref.\cite{lhaaso_nature}. AS$\gamma$'s data is taken from Ref.\cite{ASg_2228}. HAWC's data is taken from Ref.\cite{hawc_g106}. VERITAS's data is taken from Ref.\cite{VERITAS_2228}. MAGIC's data is taken from Ref.\cite{Magic2023_g106}. Fermi-LAT's data is taken from Ref.\cite{Xin2019} and Ref.\cite{Liu20}. The solid and dashed curves show for reference the expected flux with the simple one-zone model in the hadronic scenario and the leptonic scenario, respectively, given by Ref.\cite{Ge21}.  }
    \label{fig:G106}
\end{figure}

This SNR-PWN complex has been detected in the \gray band, from $\sim$1 GeV to multi-hundred TeV  by various instruments \citep{lhaaso_nature, ASg_2228, hawc_g106, magic_g106, Milagro2007, VERITAS_2228, Xin2019} as shown in Fig.~\ref{fig:G106}. The Fermi-LAT observations detected an extended source with radius $0.25^\circ$ in the tail region, which is in spatial coincidence with a molecular cloud\citep{Xin2019, Fangke2022}. Such an association is corroborated by the observation of AS$\gamma$ in $6-115\,$TeV \citep{ASg_2228}. LHAASO's measurement \citep{lhaaso_nature} at the UHE energy band shows a source centroid consistent with the position of the SNR tail while the spatial extension of the source also covers the head region. The spectrum massured by LHAASO extends up to 500\,TeV without an obvious cutoff feature.  A simple one-zone leptonic model cannot explain the broadband \gray spectrum because the IC radiation of electrons at high energies is suppressed by the KN effect. Both the spectral and the morphological measurements  seem to favor a hadronic origin of the \gray emission in the SNR tail and the existence of a proton PeVatron in this region. 

The most plausible candidate of the PeVatron is the SNR shock, from which accelerated protons may escape and illuminate the molecular cloud \citep{Bao21}. Nonthermal X-ray emission is discovered from the tail region \citep{Ge21, Fujita21}, which is emitted by electrons accelerated {\em in situ} according to the X-ray intensity profile \citep{Ge21}. It indicates a high shock velocity of at least several thousands of km/s presented in the tail region \citep{zirakashvili07}, making acceleration of PeV protons from the SNR shock available. If the Boomerang Nebula and SNR G106.3+2.7 are truly associated, the high shock velocity makes the SNR quite unusual given its age inferred from the pulsar. It is speculated that the shock in the tail direction has not been decelerated since it is expanding in a low-density cavity \citep{Ge21}, which may be created by the stellar wind or supernova explosion of previous generations of stars \citep{Kothes01}. In other words, such a special environment makes its shock maintained a high speed for a long time, which is in favor of acceleration of PeV protons. %G106.3+2.7 might represent a peculiar kind of SNRs, i.e., middle-aged SNRs in low-density cavity, which constitute another population of PeVatron candidates.

\section{Pulsar Halos}
Pulsar halos are believed to be formed by the pulsars alone with the associated SNRs disappeared due to either proper motions of the pulsars out of the SNRs or fading away of the SNRs \citep{Giacinti20}. This generally begins at $\sim 100$\,kyr after the birth of a pulsar. Pulsar halos have been discovered to be VHE \gray sources with spectra likely extending to the UHE regime. In this section, we briefly introduce the observations of HAWC and LHAASO, as well as the underlying physics of pulsar halos. Readers can refer to recent reviews \citep{Liu22,Lopez22,Fang22} for detailed discussions. 

%\subsection{Observations of Pulsar Halos around Geminga, Monogem and PSR~J0622+3749}
%\subsection{Discovery of TeV Pulsar Halos}
Discovery of extended multi-TeV \gray emission around the Geminga pulsar (PSR~J0633+1746) and the Monogem pulsar (PSR~J0659+1414) by HAWC \citep{hawc_geminga} indicates that the PWNe of these two middle-aged pulsars still remain to be efficient particle accelerators and inject a considerable amount of ultrarelativistic e$^+$e$^-$ pairs into the ambient ISM. Similar to PWNe, the spatially extended \gray emissions of pulsar halos are also produced by IC radiation of electron/positron pairs having escaped to ISM. This is different from PWNe where e$^+$e$^-$ pairs are confined in PWNe, much smaller regions than halos. The spectra of pulsar halos measured by HAWC continue up to 40\,TeV without clear cutoff features. This indicates injection of pairs with energies $>100\,$TeV  from the pulsars. The sizes of sources are at least 20-30\,pc which is about two orders of magnitude larger than the typical size of a bow-shock PWN. Hence, they are regarded as a new category  of \gray sources and termed as pulsar halos or TeV halos. Intriguingly, the steep declining profiles of the surface brightness with the distance from the pulsars measured by HAWC suggest a diffusion coefficient of particles inside the halos 2-3 orders of magnitude lower than the average diffusion coefficient in ISM derived from measurements of the ratio between secondary and primary CRs \citep{Amato2018}. The origin of such a slow diffusion is still unclear and under debate.

Given the pulsars in Geminga and Monogem not being special, one may expect existence of halos around other middle-aged pulsars \citep{Linden17, Sudoh19}.
HAWC and LHAASO have detected some extended sources in spatial association with energetic pulsars of comparable ages  to those in Geminga and Monogem. However, many of them cannot be unambiguously identified as pulsar halos yet. 
Among them, the most promising one is the extended source LHAASO~J0621+3755 \citep{LHAASO_0621}, where a middle-aged pulsar J0622+3749 is located at the center of the source. The pulsar has the comparable characteristic age, rotation period, and spin-down power, i.e, 208\,kyr, 0.333\,s and $2.7\times 10^{34}\rm erg/s$, respectively, with the Geminga pulsar (342\,kyr, 0.237\,s, and $3.3\times 10^{34}\rm erg/s$) and the Monogem pulsar (110\,kyr, 0.385\,s, and $3.8\times 10^{34}\rm erg/s$). No other plausible astrophysical counterpart if found in the region around this source. Fitting the morphology with a 2-dimensional  Gaussian template, a radius of $0.6^\circ$ is found to  contain 68\% photon flux from the source, corresponding to a spatial size could be 17\,pc according to the distance of the pulsar about 1.6\,kpc. It is worth noting that the distance is estimated based on the correlation between the \gray luminosity and the spin-down power of \gray pulsars \citep{SaZPark2010}. The physical size is comparable to the halos of Geminga and Monogem, although the angular size is much smaller. However, current accumulation of data does not support a statistically significant claim of the identification.    One may have to wait for a couple of years to have a decisive conclusion on this source. 

One of the key issues to understand pulsar halos is the origin of the slow diffusion of injected  e$^+$e$^-$ pairs. An intuitive interpretation is the existence of a highly turbulent interstellar magnetic field around those middle-aged pulsars. The strong turbulence could be either extrinsically driven at small scales \citep{Lopez18} or self-generated by particle themselves via the streaming instability \citep{Evoli18, Mukhopadhyay21}. 
%The latter mechanism may not work for those high-energy CR pairs because of insufficient injection rate\citep{Fang19}.  
Alternatively, a low-level turbulence scenario may also explain the slow diffusion given a small inclination angle between the average magnetic field direction and the observer's line of sight \citep{liu19ani}. In this case, the required slow diffusion can be ascribed to the cross-field diffusion of CRs which is largely suppressed. So far, the consensus on the origin of the slow diffusion has not been reached \citep{Fang19, Yan22, Luque22}. It is suggested that the operation of LHAASO for several years would be able to distinguish different scenarios \citep{Yan22}. On the other hand, the multi-wavelength observations combining those in GeV $\gamma$-ray band\citep{Xi19, Dimauro19} and X-ray band \citep{liu19} are also helpful to understand the nature of pulsar halos.

\section{Diffuse UHE Gamma-ray Emission from the Galactic Plane}
Galactic CRs are expected to be accelerated by sources in the Galactic Plane (GP). The average gas density in GP are also believed to be much higher than in the Galactic halo. Thus, both the higher CR intensity and the gas density predict the GP should be a bright $\gamma$-ray emitter. Indeed the bright diffuse $\gamma$-ray emission in GP is one of the most prominent feature in GeV $\gamma$-ray sky \citep{fermi_diffuse_old}. At higher energies, the diffuse emission is  also detected by the EAS arrays, Milargro \citep{milagro_diffuse} and ARGO-YBJ \citep{argo_diffuse}. The Milagro measurement extends the SEDs of diffuse $\gamma$-ray emissions to $\sim15~\rm TeV$, and the flux are in consistency with the prediction by GALPROP code. H.E.S.S also detected the diffuse $\gamma$-ray emission around $1~\rm TeV$ \citep{hess_diffuse}. However, due to its limited field of view and the background subtraction method,  H.E.S.S can hardly resolve the large scale variation of the diffuse emissions such as the Galactic IC emission. 
%The H.E.S.S team interpreted the detected diffuse emission as a  mix of diffuse Galactic $\gamma$-ray emission and unresolved sources. 

Galactic diffuse $\gamma$-ray emissions are regarded as an important tool to trace the propagation of Galactic CRs. For PeV CRs, the diffuse $\gamma$-ray emissions above 100~TeV are crucial. Recently, the AS$\gamma$ experiment has reported the first detection of diffuse $\gamma$-ray emissions above 100~TeV \citep{asgamma_diffuse}. Remarkably, 38 $\gamma$-like events above $398~\rm TeV$ are detected in GP without association with any known source. This may indicate the existence of CRs beyond  few PeV in GP.  The measured $\gamma$-ray flux above $398~\rm TeV$ are slightly higher than the prediction by models such as Galprop \citep{asgamma_diffuse}. However, whether those photons are associated with the isolated CR sources are not clear giving the statistics constrained by the instrument sensitivity. Conventionally, the diffuse emission are believed to be produced by the interaction of  relatively uniform CR `sea' with gases. However, as mentioned in \citep{yang_scpma22}, the CRs escaping from the sources can produced very extended $\gamma$-ray emissions. In this regard, although the AS$\gamma$ detected UHE photons are far from known TeV sources, there still is probability that they are from sources not resolvable by the AS$\gamma$ detector. 

%One example is the Cygnus region in which 4 UHE photons are detected by AS$\gamma$ indeed. However, the flux is still in consistence with the even higher flux observed by Milagro \citep{milagro_diffuse} within the statistic uncertainties.
%and latest LHAASO detection of giant CR bubbles in this region. It is also possible that the other photons detected by As$\gamma$ are also from similar, but weaker structures. As calculated in  \citep{yang_scpma22} for such extended structures, if they are dim, the first several photons collected by the detectors are more likely far from the CR source. In this regard, it is possible that As$\gamma$ detected diffuse $\gamma$-ray emissions are not from the CR "sea" interacting with ambient gas, but from CR "islands", which is the high CR density regions near the CR sources. 

The straight forward way to pin down such ambiguity is either resolving those sources with more sensitive detectors, such as LHAASO, or improving the measurements of diffuse emission in the entire UHE domain. This requires not only much significant detection of diffuse emission, but also   distinguish between photons from the `true' diffuse emission and from the discrete faint sources. Recently finished analyses of LHAASO observations on both the UHE diffuse $\gamma$-ray distribution in the northern sky\cite{cao2023-diffuse-gamma} and the catalog of UHE sources \cite{lhaaso_catalog} have taken the first step towards the goal of the precise measurements.

\section{Summary}
By discovering more than a dozen UHE gamma-ray sources, LHAASO has thoroughly opened the window of UHE gamma-ray astronomy. Those sources reveal that our Galaxy is full of powerful particle accelerators known as PeVatrons, thus shed light on the puzzle of the origin of cosmic rays. The possible astrophysical counterparts of the PeVatrons are diverse including pulsar wind nebulae, supernova remnants, and star-forming regions. This not only largely enriches the fascinating UHE astronomy, but also strongly implies that CRs are sources from various types of factories. 

Amongst the sources, the Crab nebula is the only one that has been firmly localized and identified. Investigation in-depth with the hundreds of UHE photons discovers that the Crab nebula is an extreme electron PeVatron accelerating particles at a rate close to the theoretical limit. Moreover, the SED in UHE band, around 1 PeV in particular, indicates a deviation from the standard one-zone leptonic model and a hint of a hadronic component.

In conclusion, UHE  $\gamma$-ray astronomy opens a wide field for further exploration of new radiation mechanisms and, more importantly, exploration of CR particle acceleration and propagation
within source regions. Observation of diffuse $\gamma$-ray distribution will provide essential information about the transportation of the CRs in our Galaxy, which is related to the origin of the knee structure of the CR spectrum. Furthermore, UHE $\gamma$-ray observation opens up a new energy domain for indirect searches of dark matter  \citep{lhaaso_dm} and tests of fundamental physics laws \citep{lhaaso_liv}, which will help us explore potential new physics in unprecedented parameter spaces. This review was motivated by reviewing the status of the completely new field of UHE $\gamma$-ray astronomy, but it will likely raise a series of questions to be addressed in future investigations.

%Disclosure
\section*{DISCLOSURE STATEMENT}

% Acknowledgements
\section*{ACKNOWLEDGMENTS}
This work is funded by the National Key R$\&$D program of China under the grants 2018YFA0404204 and 2018YFA0404201, the Chengdu Management Committee of Tianfu New Area,  the NSFC under the grant 12022502, U2031105.
Authors also appreciate the proof reading and efforts in improving English presentation of the manuscript by Andrew J Cao.

% References
%
% Margin notes within bibliography

\bibliography{ref}
\bibliographystyle{ar-style5}

\end{document}